\pdfoutput=1
\documentclass[prl,twocolumn,showpacs,amsmath,amssymb,floatfix,superscriptaddress]{revtex4-1}

\usepackage{graphicx}
\usepackage{dcolumn}
\usepackage{bm}
\usepackage{hyperref}

\begin{document}

\title{Proximity band structure and spin textures on both sides of topological-insulator/ferromagnetic-metal interface and their transport probes}

\author{J. M. Marmolejo-Tejada}
\affiliation{Department of Physics and Astronomy, University of Delaware, Newark, DE 19716-2570, USA}
\affiliation{School of Electrical and Electronics Engineering, Universidad del Valle, Cali, AA 25360, Colombia}
\author{K. Dolui}
\affiliation{Department of Physics and Astronomy, University of Delaware, Newark, DE 19716-2570, USA}
\author{P. Lazi\'{c}}
\affiliation{Rudjer Bo\v{s}kovi\'{c} Institute, PO Box 180, Bijeni\v{c}ka c. 54, 10 002 Zagreb, Croatia}
\author{P.-H. Chang}
\affiliation{Department of Physics and Astronomy, University of Nebraska Lincoln, Lincoln, Nebraska 68588, USA}
\author{S. Smidstrup}
\affiliation{QuantumWise A/S, Fruebjergvej 3, Box 4, DK-2100 Copenhagen, Denmark}
\author{D. Stradi}
\affiliation{QuantumWise A/S, Fruebjergvej 3, Box 4, DK-2100 Copenhagen, Denmark}
\author{K. Stokbro}
\affiliation{QuantumWise A/S, Fruebjergvej 3, Box 4, DK-2100 Copenhagen, Denmark}
\author{B. K. Nikoli\'{c}}
\email{bnikolic@udel.edu}
\affiliation{Department of Physics and Astronomy, University of Delaware, Newark, DE 19716-2570, USA}

\begin{abstract}
The control of recently observed spintronic effects in topological-insulator/ferromagnetic-metal (TI/FM) heterostructures is thwarted by the lack of understanding of band structure and spin texture around their interfaces. Here we combine density functional theory with Green's function techniques to obtain the spectral function at any plane  passing through atoms of Bi$_2$Se$_3$ and Co or Cu layers comprising the interface. In contrast to widely assumed but thinly tested Dirac cone gapped by the proximity exchange field, we find that the Rashba ferromagnetic model describes  the spectral function on the surface of Bi$_2$Se$_3$ in contact with Co near the Fermi level $E_F^0$, where circular and snowflake-like constant energy contours coexist around which spin locks to momentum. The remnant of the Dirac cone  is hybridized with evanescent wave functions injected by metallic layers and pushed, due to charge transfer from Co or Cu layers, few tenths of eV below $E_F^0$ for both Bi$_2$Se$_3$/Co and Bi$_2$Se$_3$/Cu interfaces while hosting distorted helical spin texture wounding around a single circle. These features explain recent observation [K. Kondou {\em et al.}, Nat. Phys. {\bf 12}, 1027 (2016)] of sensitivity of spin-to-charge conversion signal at TI/Cu interface to tuning of $E_F^0$. Interestingly, three monolayers of Co adjacent to Bi$_2$Se$_3$ host spectral functions  very different from the bulk metal, as well as in-plane spin textures signifying the spin-orbit proximity effect.  We predict that out-of-plane tunneling anisotropic magnetoresistance in vertical heterostructure Cu/Bi$_2$Se$_3$/Co, where current flowing perpendicular to its interfaces is modulated by rotating magnetization from parallel to orthogonal to current flow, can serve as a sensitive probe 
of spin texture residing at $E_F^0$. 
\end{abstract}

\maketitle

The  recent experiments on spin-orbit torque (SOT)~\cite{Mellnik2014,Fan2016} and spin-to-charge conversion~\cite{Shiomi2014,Kondou2016} in topological-insulator/ferromagnetic-metal (TI/FM) heterostructures have ignited the field of {\em topological spintronics}. In these devices, giant non-equilibrium spin densities~\cite{Pesin2012,Chang2015,Mahfouzi2014a,Mahfouzi2016} are expected to be generated due to strong spin-orbit coupling (SOC) on metallic surfaces of three-dimensional (3D) TIs and the corresponding (nearly~\cite{Bansil2016}) helical spin-momentum locking along a single Fermi circle for Dirac electrons hosted by those surfaces~\cite{Hasan2010}. Such strong interfacial SOC-driven phenomena are also envisaged to underlie a plethora of novel spintronic technologies~\cite{Soumyanarayanan2016}. 

\begin{figure}
	\includegraphics[scale=0.45,angle=0]{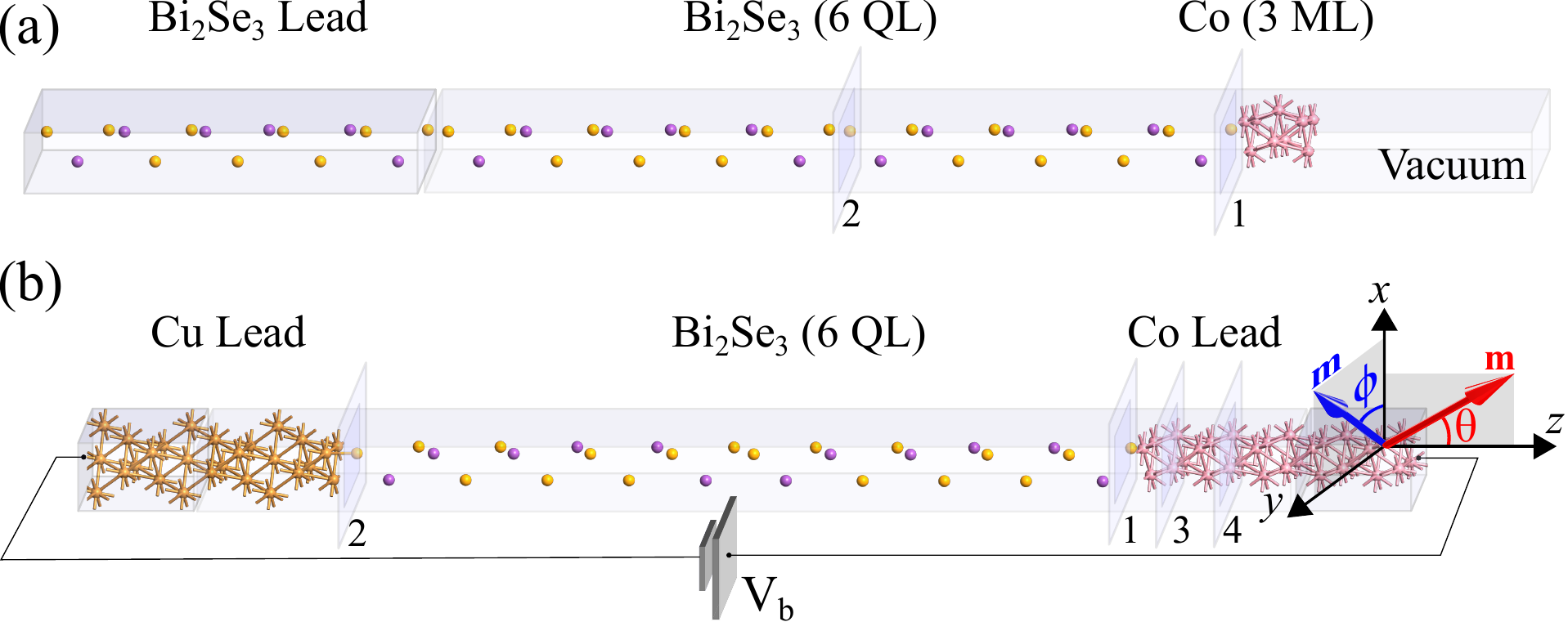}
	\caption{Schematic view of TI-based heterostructures where: (a) semi-infinite Bi$_2$Se$_3$ layer is attached to $n$ monolayers of Co(0001); (b) 6 QLs of Bi$_2$Se$_3$ are sandwiched between  semi-infinite Cu(111) layer and semi-infinite Co(0001) layer. Both heterostructures are infinite in the transverse direction, so that the depicted supercells are periodically repeated within the $xy$-plane. The magnetization $\mathbf{m}$ of the Co layer is fixed along the $z$-axis in (a), or rotated within the $xy$-plane or the $xz$-plane in (b). Applying the bias voltage $V_b$ to the {\em vertical} heterostructure in panel (b) leads to a charge current flowing {\em perpendicularly} to both Bi$_2$Se$_3$/Cu and Bi$_2$Se$_3$/Co interfaces.}
	\label{fig:fig1}
\end{figure}

\begin{figure*}
	\includegraphics[scale=0.28,angle=0]{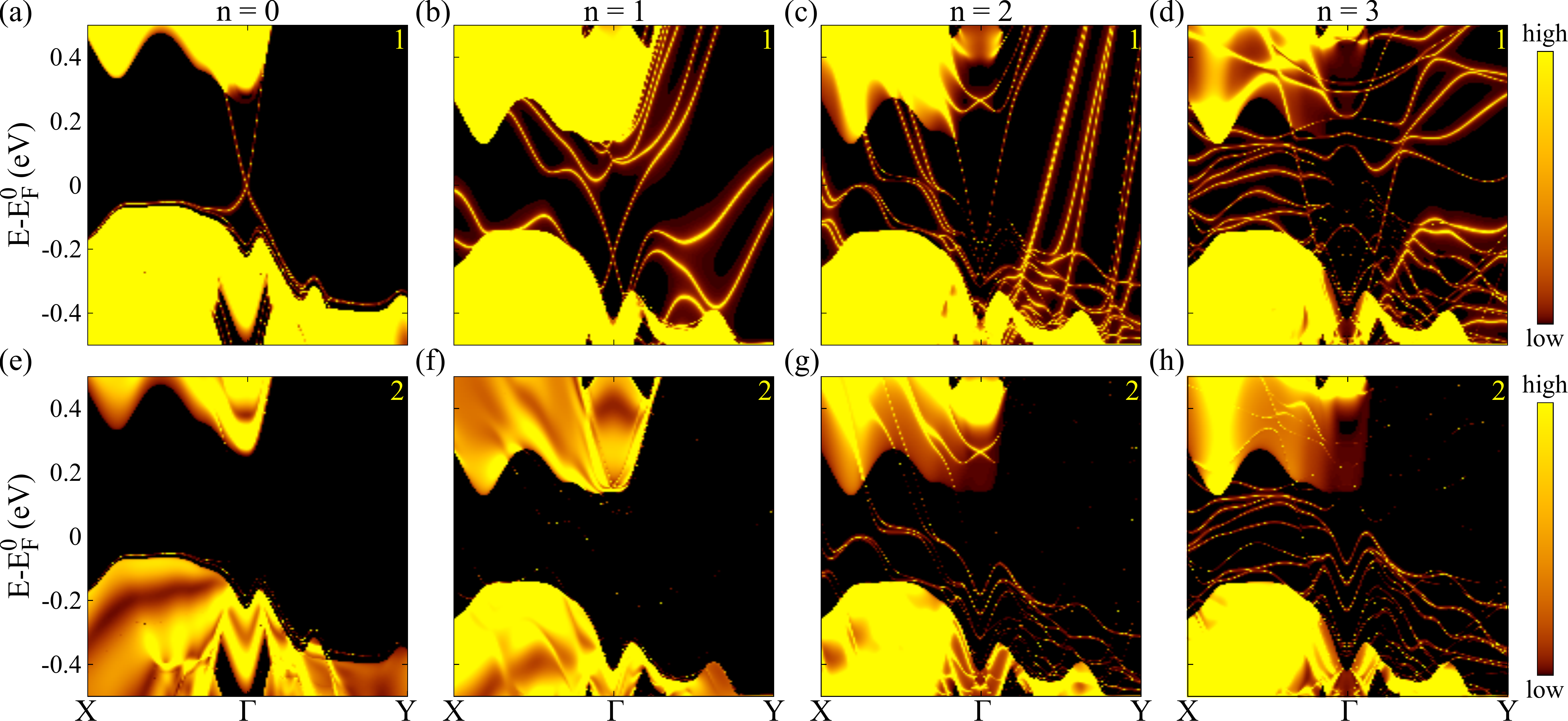}
	\caption{Spectral function, defined in Eq.~\eqref{eq:spectral}, at plane $1$ for panels (b)--(d) or plane $2$ for panels (f)--(h) within Bi$_2$Se$_3$/Co($n$ ML) heterostructure in Fig.~\ref{fig:fig1}(a) with $\mathbf{m} \parallel \hat{z}$.  For comparison, panels (a) and (e) plot the spectral function at planes $1$ (akin to Ref.~\cite{Zhang2009}) and $2$, respectively, within semi-infinite Bi$_2$Se$_3$ crystal in contact with vacuum (i.e., $n=0$). From $\Gamma$ to $Y$ we plot $A(E;k_x=0,k_y;z \in \{1,2\})$, while from $\Gamma$ to $X$ we plot $A(E;k_x,k_y=0;z \in \{1,2\})$.}
	\label{fig:fig2}
\end{figure*}

These effects have been interpreted almost exclusively using simplistic models, such as the Dirac Hamiltonian for the TI surface with an additional Zeeman term describing coupling of magnetization of the FM layer to the surface state spins~\cite{Hasan2010}, \mbox{$\hat{H}^\mathrm{Dirac} = v_F (\hat{\bm \sigma} \times \hat{\mathbf{p}})_z - \Delta \mathbf{m} \cdot \hat{\bm \sigma}$}, where $\hat{\mathbf{p}}$ is the momentum operator, $\hat{\bm \sigma}$ is the vector of the Pauli matrices, $\mathbf{m}$ is the magnetization unit vector and $v_F$ is the Fermi velocity. Thus, the only effect of FM layer captured by  $\hat{H}^\mathrm{Dirac}$ is proximity effect-induced exchange coupling  $\Delta$ which opens a gap in the Dirac cone energy-momentum dispersion~\cite{Hasan2010}, thereby making Dirac electrons massive. On the other hand, recent first-principles calculations~\cite{Luo2013,Lee2014a} demonstrate that band structure of even TI/ferromagnetic-insulator (TI/FI) bilayers, where hybridization between TI and FI states is largely absent, cannot be captured by simplistic models like $\hat{H}^\mathrm{Dirac}$. The properties of TI/FM interfaces are far more complex due to injection of evanescent wave functions from the FM layer into the bulk gap of the TI layer, which can hybridize with surface state of TI and blur its Dirac cone (as already observed in tight-binding models of TI/FM interfaces~\cite{Mahfouzi2014a,Zhao2010}), as well as related charge transfer. Thus, the {\em key issue} for topological  spintronics~\cite{Mellnik2014,Fan2016,Shiomi2014,Kondou2016,Soumyanarayanan2016} is to understand band structure and spin textures (including the fate of the Dirac cone and its spin-momentum locking)  in hybridized TI with FM or normal metal (NM)~\cite{Kondou2016} layers  at nanometer scale around the interface where they are brought into contact, where properties of both TI side and FM or NM side of the interface can be quite different from the properties of corresponding bulk materials. 

For example, computational searches~\cite{Zhang2009} for new materials realizing 3D TIs (or other topologically nontrivial electronic phases of matter like Weyl semi-metals~\cite{Soluyanov2015} and Chern insulators~\cite{Sheng2016}) have crucially relied on first-principles calculations of spectral function on their boundaries and its confirmation by  spin- and angle-resolved photoemission spectroscopy (spin-ARPES)~\cite{Shoman2015}. A  standard density functional theory (DFT)-based framework developed for this purpose---where DFT band structure around the Fermi level $E_F^0$ is reconstructed using the Wannier tight-binding Hamiltonian~\cite{Marzari2012} used to obtain the retarded Green's function (GF) of semi-infinite homogeneous crystal and the spectral function on its surface in contact with vacuum~\cite{Zhang2009,Soluyanov2015,Sheng2016}---is difficult to apply to complicated inhomogeneous systems like TI/FM bilayers due to strongly entangled bands in the region of interest around $E_F^0$. Also, spin-ARPES experiments cannot probe buried interfaces below too many monolayers (e.g., penetration depth of low-energy photons is 2--4 nm) of FM or NM deposited onto the TI surface~\cite{Shoman2015}.

\begin{figure*}
	\includegraphics[scale=0.255,angle=0]{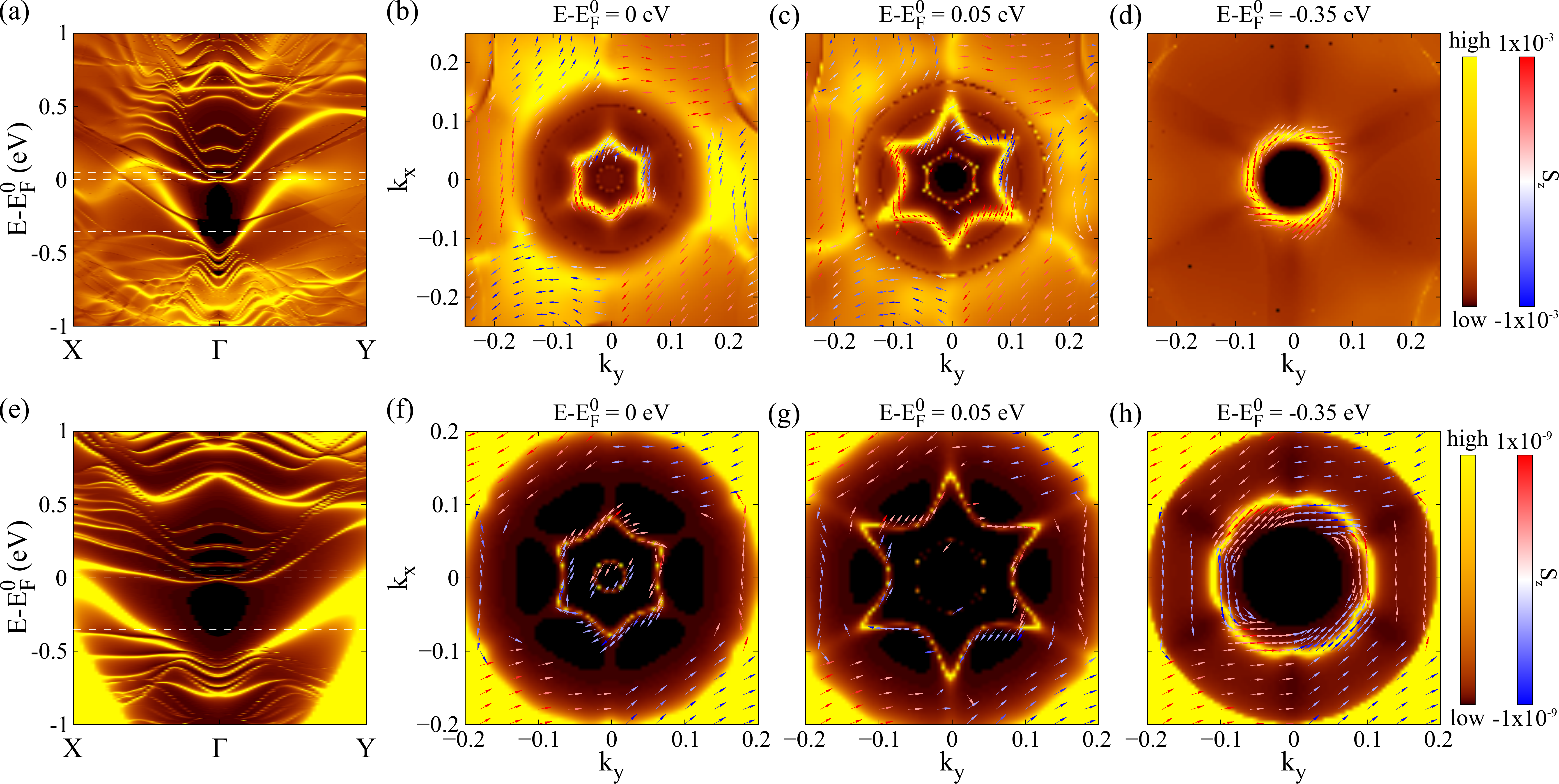}
	\caption{Spectral function at: (a)--(d) plane $1$ in Fig.~\ref{fig:fig1}(b) which is passing through Se atoms at the Bi$_2$Se$_3$/Co interface with \mbox{$\mathbf{m} \parallel \hat{z}$}; and (e)--(h) plane $2$ in Fig.~\ref{fig:fig1}(b) which is passing through Se atoms at Bi$_2$Se$_3$/Cu interface, where we remove Co layer to make Bi$_2$Se$_3$ semi-infinite along the $z$-axis. In panels (a) and (e), we plot  $A(E;k_x=0,k_y;z \in \{1,2\})$ from $\Gamma$ to $Y$ and $A(E;k_x,k_y=0;z \in \{1,2\})$ from $\Gamma$ to $X$. Panels (b)--(d) and (f)--(h) plot constant energy contours of \mbox{$A(E-E_F^0 \in \{0.0 \ \mathrm{eV},0.05  \ \mathrm{eV},-0.35  \ \mathrm{eV}\};k_x,k_y;z \in \{1,2\})$}  at three energies marked by horizontal dashed lines in panels  (a) or (e), respectively, as well as the corresponding spin textures where the out-of-plane $S_z$ component is indicted in color (red for positive and blue for negative).  The units for $k_x$ and $k_y$ are $2\pi/a$ where $a$ is the lattice constant of a common supercell combining two unit cells of the two layers around the corresponding interface.}
	\label{fig:fig3}
\end{figure*}

An attempt~\cite{Spataru2014} to obtain the spectral function, $A_j(E;{\mathbf k})=\sum_{n,i}^{i \in \mathrm{QL}_j} w_{n\mathbf{k}}^i \delta(E-\varepsilon_{n\mathbf{k}})$, directly from DFT computed energy-momentum dispersion $\varepsilon_{n\mathbf{k}}$ ($n$ is the band index and  $\mathbf{k}$ is the crystal momentum) and site-projected  character $w_{n\mathbf{k}}^i$ of the corresponding eigenfunctions for TI/FM supercells has produced ambiguous results. This is due to arbitrariness in broadening the delta function $\delta(E-\varepsilon_{n\mathbf{k}})$, as well as due to  usage of atomic sites $i$ within the whole $j$ quintuple layer (QL$_j$) of Bi$_2$Se$_3$ (one QL consists of three Se layers strongly bonded to two Bi layers in between)   which effectively averages the spectral function over all geometric planes within QL$_j$. Similar ambiguities (such as setting the amount of electron density which is localized on the surface or within the whole interfacial QL) plague interpretation of projected DFT band structure of TI/FI~\cite{Lee2014a} and TI/FM bilayers~\cite{Zhang2016}.

Here we develop a framework which combines the noncollinear DFT Hamiltonian $\mathbf{H}^\mathrm{DFT}$,  represented in a basis of variationally optimized localized atomic orbitals~\cite{Ozaki2003}, with retarded GF calculations from which one can extract the spectral function and spin textures at an arbitrary geometric plane of interest within a junction combining TI, FM and NM layers. It also makes it possible to  compute their spin and charge transport properties in the linear-response regime or at finite bias voltage. We apply this framework to two Bi$_2$Se$_3$-based heterostructures whose supercells are depicted in Fig.~\ref{fig:fig1}, where we assume that those supercells are periodically repeated in the transverse $xy$-direction. 

The heterostructure in Fig.~\ref{fig:fig1}(a) consists of  Bi$_2$Se$_3$, chosen as the prototypical 3D TI~\cite{Bansil2016,Hasan2010},  whose surface is covered by $n$ monolayers (MLs) of Co. The retarded GF of this heterostructure is computed as
\begin{equation}\label{eq:oneleadgf}
\mathbf{G}_{\mathbf{k}_\parallel}(E)= [E-\mathbf{H}^\mathrm{DFT}_{\mathbf{k}_\parallel} - {\bm \Sigma}_{\mathbf{k}_\parallel}^{\mathrm{Bi_2Se_3}}(E)]^{-1},
\end{equation}
where $\mathbf{k}_\parallel=(k_x,k_y)$ is the transverse $k$-vector, ${\bm \Sigma}_{\mathbf{k}_\parallel}^{\mathrm{Bi_2Se_3}}(E)$ is the self-energy~\cite{Velev2004} describing the semi-infinite Bi$_2$Se$_3$ lead and $\mathbf{H}^\mathrm{DFT}_{\mathbf{k}_\parallel}$ is the Hamiltonian of the {\em active region} consisting of  $n$ MLs of cobalt plus 6 QLs of Bi$_2$Se$_3$ to which the lead is attached. We choose  $n=$1--3 since ultrathin FM layers of thickness \mbox{$\simeq 1$ nm} are typically employed in SOT experiments~\cite{Kim2013} in order to preserve perpendicular magnetic anisotropy (note that magnetocrystalline anisotropy does favor out-of-plane $\mathbf{m}$ in Bi$_2$Se$_3$/Co bilayers~\cite{Zhang2016}). The spectral function (or local density of states) at an arbitrary plane at position $z$ within the active region is computed from
\begin{equation}\label{eq:spectral}
A(E;k_x,k_y,z)=-\mathrm{Im}\,[G_{\mathbf{k}_\parallel}(E;z,z)]/\pi,
\end{equation}
where the diagonal matrix elements $G_{\mathbf{k}_\parallel}(E;z,z)$ are obtained by transforming Eq.~\eqref{eq:oneleadgf} from orbital to a real-space representation.

\begin{figure*}
	\includegraphics[scale=0.255,angle=0]{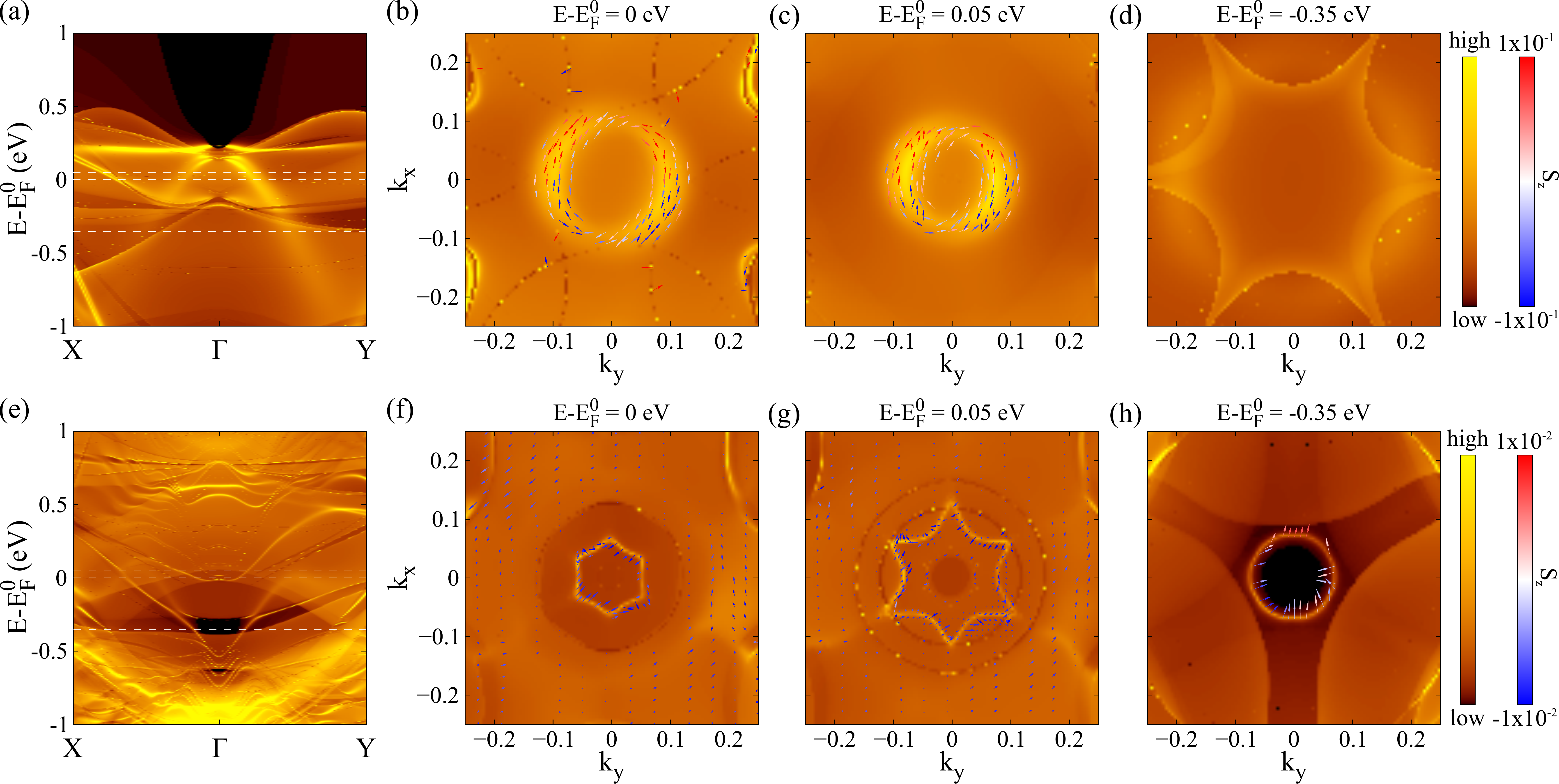}
	\caption{Spectral function at: (a)--(d) the surface of a semi-infinite Co layer in contact with vacuum where $\mathbf{m}$ is perpendicular to the surface; and (e)--(h) plane 3 in Fig.~\ref{fig:fig1}(b) which is passing through Co atoms at the Bi$_2$Se$_3$/Co interface with $\mathbf{m} \parallel \hat{z}$. In panel (e), we plot  $A(E;k_x=0,k_y;z \in 3)$ from $\Gamma$ to $Y$ and $A(E;k_x,k_y=0;z \in 3)$ from $\Gamma$ to $X$. Panels (b)--(d) and (f)--(h) plot constant energy contours of the spectral function at three energies marked by horizontal dashed lines in panels (a) or (e), respectively, as well as the corresponding spin textures where the magnitude of the out-of-plane $S_z$ component is indicted in color (red for positive and blue for negative).  The units for $k_x$ and $k_y$ are $2\pi/a$ where $a$ is the lattice constant of a common supercell combining two unit cells of the two layers around the corresponding interface.}
	\label{fig:fig3A}
\end{figure*}

The  heterostructure in Fig.~\ref{fig:fig1}(b) consists of semi-infinite Cu and Co leads sandwiching a Bi$_2$Se$_3$ layer of finite thickness, where we choose Cu as the NM layer similar to the very recent spin-to-charge conversion experiment of Ref.~\cite{Kondou2016}. Such a heterostructure is termed {\em vertical}  or {\em current-perpendicular-to-plane} in spintronics terminology since applying bias voltage $V_b$ drives a current perpendicularly to the TI/FM interface. Its retarded GF is computed as
\begin{equation}\label{eq:twoleadgf}
\mathbf{G}_{\mathbf{k}_\parallel}(E) = [E-\mathbf{H}^\mathrm{DFT}_{\mathbf{k}_\parallel} - {\bm \Sigma}_{\mathbf{k}_\parallel}^{\mathrm{Cu}}(E) - {\bm \Sigma}_{\mathbf{k}_\parallel}^{\mathrm{Co}}(E)]^{-1},
\end{equation}
where $\mathbf{H}^\mathrm{DFT}_{\mathbf{k}_\parallel}$ describes the active region consisting of 6 QLs of Bi$_2$Se$_3$ plus $4$ MLs of Cu and $4$ MLs of Cu. Its linear-response resistance $R$ is given by the Landauer formula
\begin{equation}\label{eq:resistance}
\frac{1}{R} = \frac{e^2}{h \Omega_\mathrm{BZ}} \int_\mathrm{BZ} \!\! d\mathbf{k}_\parallel \int \!\! dE\, \left(-\frac{\partial f}{\partial E} \right) 
\mathrm{Tr}[{\bm \Gamma}_{\mathbf{k}_\parallel}^\mathrm{Co} \mathbf{G}_{\mathbf{k}_\parallel} {\bm \Gamma}_{\mathbf{k}_\parallel}^\mathrm{Cu} \mathbf{G}_{\mathbf{k}_\parallel}^\dagger],
\end{equation}
where we assume temperature $T=300$ K in the Fermi-Dirac distribution function $f(E)$, ${\bm \Gamma}_{\mathbf{k}_\parallel}^\alpha =i({\bm \Sigma}_{\mathbf{k}_\parallel}^{\alpha} - [{\bm \Sigma}_{\mathbf{k}_\parallel}^{\alpha}]^\dagger)$ and $\Omega_\mathrm{BZ}$ is the area of the two-dimensional (2D) Brillouin zone (BZ) within which $\mathbf{k}_\parallel$ vectors are sampled.

The spectral function of the heterostructure in Fig.~\ref{fig:fig1}(a) computed at planes 1 and 2 within the Bi$_2$Se$_3$ layer is shown in Figs.~\ref{fig:fig2}(b)--(d) and ~\ref{fig:fig2}(f)--(h), respectively, where plane 1 is passing through Se atoms on the  Bi$_2$Se$_3$ surface in contact with Co layer and plane 2 is three QLs (or \mbox{$\simeq 2.85$ nm}) away from plane 1. For comparison, we also show in Figs.~\ref{fig:fig2}(a) and ~\ref{fig:fig2}(b) the spectral function at the same two planes within the semi-infinite Bi$_2$Se$_3$ layer in contact with vacuum, thereby reproducing the results from Ref.~\cite{Zhang2009} by our formalism. While the Dirac cone at the $\Gamma$-point is still intact in Fig.~\ref{fig:fig2}(b) for $n=1$ ML of Co, its Dirac point (DP) is gradually pushed into the valence band of Bi$_2$Se$_3$ with increasing $n$ because of charge transfer from metal  to TI. The charge transfer visualized in Figs.~\ref{fig:fig5}(c) and ~\ref{fig:fig5}(d) is relatively small, but due to small density of states (DOS) at the DP it is easy to push it  down until it merges with the larger DOS in the valence band of the TI. Adding more MLs of Co in Figs.~\ref{fig:fig2}(c) and ~\ref{fig:fig2}(d) also introduces additional bands within the bulk gap of Bi$_2$Se$_3$ due to injection of evanescent wave functions which hybridize with the Dirac cone. The metallic surface states of  Bi$_2$Se$_3$ itself penetrate into its bulk over a distance of 2 QLs~\cite{Chang2015}, so that in Fig.~\ref{fig:fig2}(e) the spectral function on plane 2 vanishes inside the gap of the  semi-infinite Bi$_2$Se$_3$ layer in contact with vacuum, while the remaining states inside the gap in Figs.~\ref{fig:fig2}(f)--(h) can be attributed to the Co layer.  

For infinitely many MLs of Co attached to 6 QLs of Bi$_2$Se$_3$ within the Cu/Bi$_2$Se$_3$/Co heterostructure in Fig.~\ref{fig:fig1}(b), the remnant of the Dirac cone from the TI surface can be identified in Fig.~\ref{fig:fig3}(a) at around 0.5 eV below $E_F^0$ while it is pushed even further below in the case of Cu/Bi$_2$Se$_3$ interface in Fig.~\ref{fig:fig3}(e). The difference in work functions \mbox{$\Phi_\mathrm{Co}=5.0$ eV} or \mbox{$\Phi_\mathrm{Cu}=4.7$} eV and electron affinity \mbox{$\chi_\mathrm{Bi_2Se_3}=5.3$ eV} determines~\cite{Spataru2014} the band alignment and the strength of hybridization, where $n$-type doping [see also Figs.~\ref{fig:fig5}(c) and ~\ref{fig:fig5}(d)] of the Bi$_2$Se$_3$ layer pins $E_F^0$ of the whole Cu/Bi$_2$Se$_3$/Co heterostructure in the conduction band of the bulk Bi$_2$Se$_3$. The remnant of the Dirac cone is quite different from the often assumed~\cite{Hasan2010} eigenspectrum of $\hat{H}^\mathrm{Dirac}$ because of hybridization with the valence band of Bi$_2$Se$_3$, as well as with states injected by the Co or Cu layers whose penetration into TI is visualized by plotting position- and energy-dependent spectral function $A(E;z)= \frac{1}{\Omega_\mathrm{BZ}} \int \!\! dk_x dk_y \, A(E;k_x,k_y;z)$ in Fig.~\ref{fig:fig4}(a). On the other hand, the energy-momentum dispersion in the vicinity of $E_F^0$ and for an interval of $\mathbf{k}_\parallel$ vectors around the $\Gamma$-point is surprisingly well-described by another simplistic model---ferromagnetic Rashba Hamiltonian~\cite{Nagaosa2010}. 

In Figs.~\ref{fig:fig3}(b)---(d) and Figs.~\ref{fig:fig3}(f)---(h) we show constant energy contours of the spectral function at three selected energies $E$ denoted in Figs.~\ref{fig:fig3}(a) and  ~\ref{fig:fig3}(e) by dashed horizontal lines. Instead of a single circle as the constant energy contour for the eigenspectrum of $\hat{H}^\mathrm{Dirac}$, or single hexagon or snowflake-like contours (due to hexagonal warping~\cite{Bansil2016}) sufficiently away from DP for the eigenspectrum of $\mathbf{H}^\mathrm{DFT}$ of the isolated Bi$_2$Se$_3$ layer, here we find multiple circular and snowflake-like contours close to the $\Gamma$-point. The spin textures within the constant energy contours are computed from the spin-resolved spectral function. For energies near $E=E_F^0$, the spin textures shown in Figs.~\ref{fig:fig3}(b) and ~\ref{fig:fig3}{c) are quite different from the  helical ones in isolated Bi$_2$Se$_3$ layer~\cite{Zhang2009}. Nevertheless, Fig.~\ref{fig:fig3}(d) shows that the remnant Dirac cone still generates distorted helical spin texture wounding along a single circle but with out-of-plane $S_z$ component due to the presence of Co layer. 
	
The envisaged applications of TIs in spintronics are based~\cite{Mellnik2014,Fan2016,Shiomi2014,Kondou2016,Pesin2012,Mahfouzi2014a,Mahfouzi2016} on spin textures like the one in Fig.~\ref{fig:fig3}(d) since it maximizes~\cite{Pesin2012,Chang2015} generation of nonequilibrium spin density when current is passed parallel to the TI surface. However, utilizing spin texture in Fig.~\ref{fig:fig3}(d) in lateral TI/FM heterostructures would require to shift $E_F$ (by changing the composition of TI~\cite{Kondou2016} or by applying a gate voltage~\cite{Fan2016}) by few tenths of eV below $E_F^0$ of the undoped heterostructures in Fig.~\ref{fig:fig3}(a). For example, extreme sensitivity of spin-to-charge conversion was recently observed~Ref.~\cite{Kondou2016} on the surface of (Bi$_{1-x}$Sb$_x$)$_2$Te$_3$ TI  covered by a 8 nm thick Cu layer as $E_F$ of the TI layer was tuned, which is difficult to explain by assuming that the Dirac cone on the TI surface remains intact after the deposition of the  Cu layer (e.g., Ref.~\cite{Kondou2016} had to invoke ``instability of the helical spin structure''). On the other hand, it is easy to understand from Figs.~\ref{fig:fig3}(f)--(h) interface demonstrating how spin textures at Bi$_2$Se$_3$/Cu interface change dramatically as one moves $E_F$ (even slightly) below or above  $E_F^0$. Comparing Figs.~\ref{fig:fig3}(a)--(d) with  ~\ref{fig:fig3}(e)--(h) makes it possible to understand the effect of the magnetization of the Co layer, which modifies~\cite{Nagaosa2010} Rashba dispersion around $E_F^0$ and the corresponding spin textures (particularly the out-of-plane $S_z$ component).

\begin{figure}
	\includegraphics[scale=0.21,angle=0]{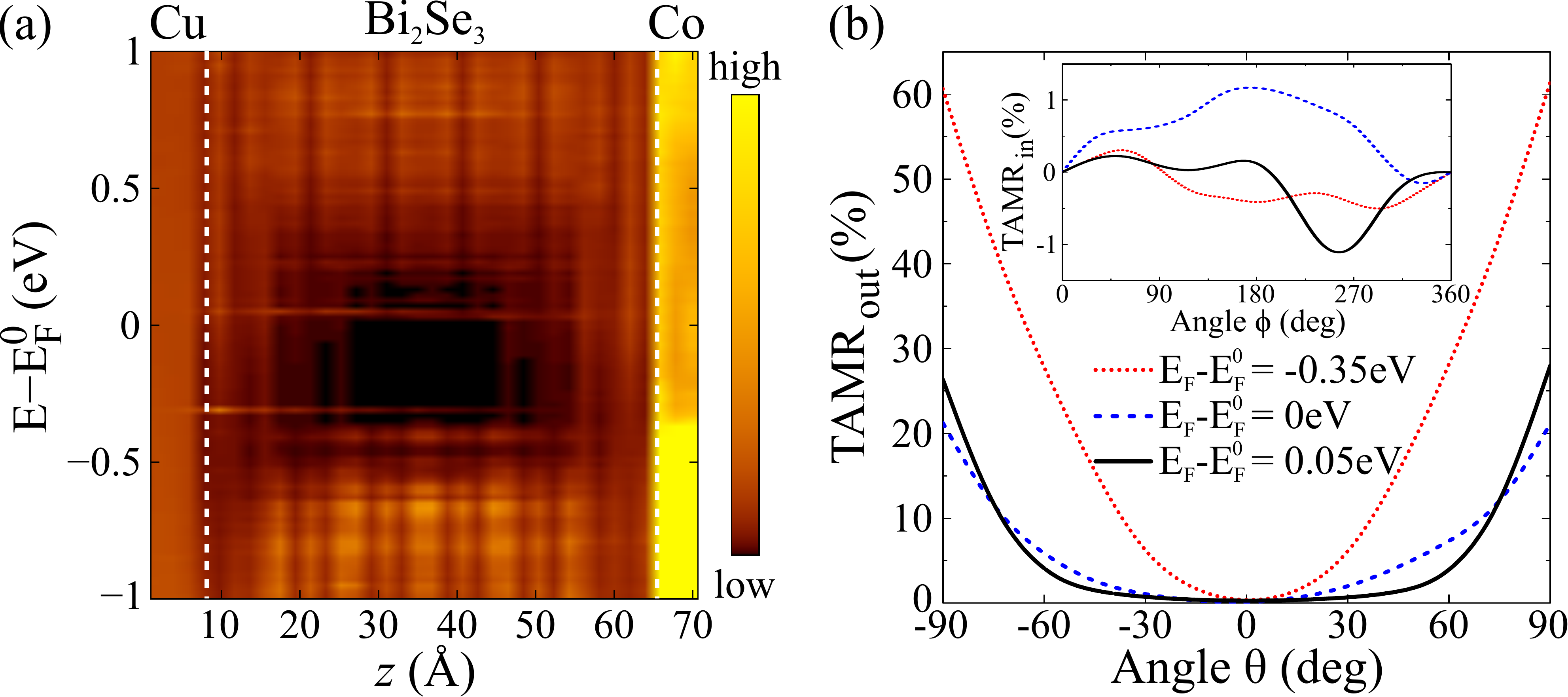}
	\caption{(a) The position- and energy-dependent spectral function $A(E;z)= \frac{1}{\Omega_\mathrm{BZ}} \int \!\! dk_x dk_y \, A(E;k_x,k_y;z)$ from the left Cu lead, across Bi$_2$Se$_3$ tunnel barrier, toward the right Co lead for the heterostructure in Fig.~\ref{fig:fig1}(b). (b) The out-of-plane $\mathrm{TAMR}_\mathrm{out}(\theta)$ ratio defined in Eq.~\eqref{eq:tamr} as function of  angle $\theta$  between the magnetization $\mathbf{m}$ and the direction of current injected along the $z$-axis in Fig.~\ref{fig:fig1}(b). Inset in panel (b) shows angular dependence of the in-plane $\mathrm{TAMR}_\mathrm{in}(\phi)$ ratio. In order to converge the integration over the transverse wave vector ${\bf k}_\parallel$ in Eq.~\eqref{eq:resistance}, we employ a uniform grid of $101 \times 101$ $k$-points for $\mathrm{TAMR}_\mathrm{out}(\theta)$ and $251 \times 251$ $k$-points for $\mathrm{TAMR}_\mathrm{in}(\phi)$.}
	\label{fig:fig4}
\end{figure}

The theoretical modeling of SOT in TI/FM~\cite{Mellnik2014,Mahfouzi2016} or heavy-metal/FM ~\cite{Haney2013} bilayers is usually conducted by starting from strictly 2D Hamiltonians, such as $\hat{H}^\mathrm{Dirac}$ or the Rashba ferromagnetic model~\cite{Nagaosa2010}, respectively, so that the FM layer is not considered explicitly. Figures~\ref{fig:fig3A}(e)--(h) show that this is not warranted since the Bi$_2$Se$_3$ layer induces proximity SOC and the corresponding in-plane spin texture on the first ML of Co, which decays to zero only after reaching plane $4$ in Fig.~\ref{fig:fig1}(b). In fact, we find non-trivial in-plane spin texture even on the surface of Co in contact with vacuum, as shown in Figs.~\ref{fig:fig3A}(b)--(d), which is nevertheless quite different from those in Figs.~\ref{fig:fig3A}(f)--(h). The in-plane spin texture in Figs.~\ref{fig:fig3A}(b)--(d) is a consequence of the Rashba SOC enabled by inversion asymmetry due to Co surface~\cite{Chantis2007} where an electrostatic potential gradient can be created by the charge distribution at the metal/vacuum interface and thereby  confine wave functions into a Rashba spin-split quasi-2D electron gas~\cite{Bahramy2012}. Thus,  Figs.~\ref{fig:fig3A}(a)--(d) explains the  origin of recently observed~\cite{Emori2016} SOT in the absence of any adjacent heavy metal or TI layer since passing current parallel to MLs of Co hosting nonzero in-plane spin textures will generate a nonequilibrium spin density~\cite{Chang2015} $\mathbf{S}_\mathrm{neq}$ and spin-orbit torque $\propto \mathbf{S}_\mathrm{neq}  \times \mathbf{m}$~\cite{Mahfouzi2016,Haney2013}.  

\begin{figure*}
	\includegraphics[scale=0.2,angle=0]{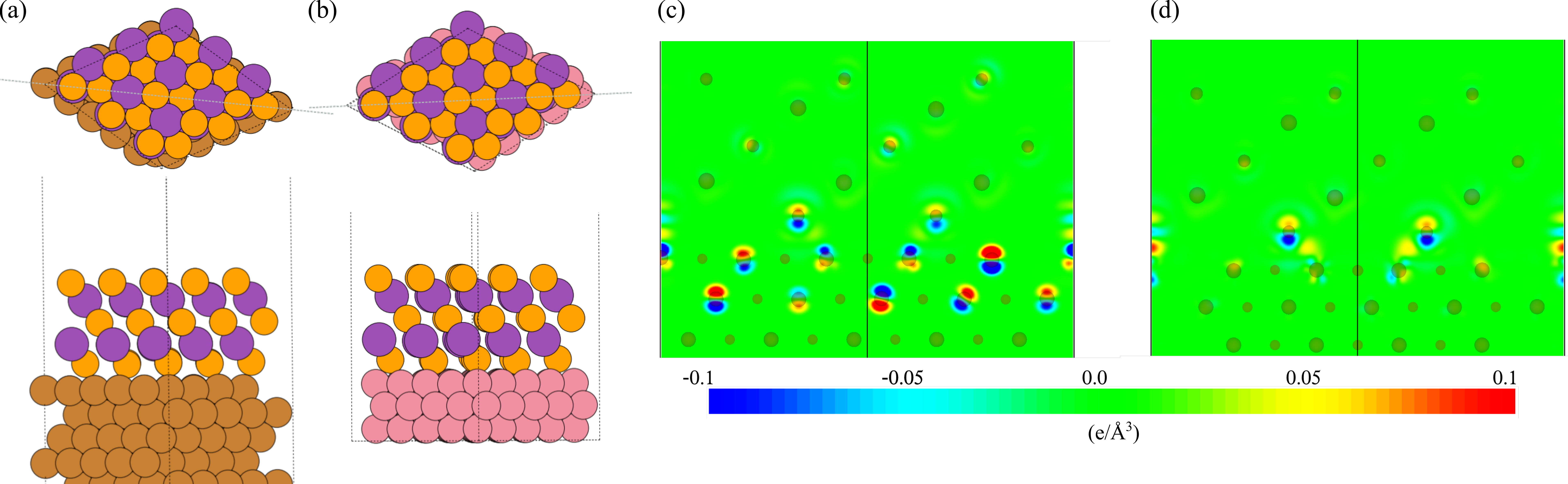}
	\caption{(Color online). Top and side view of common unit cells for (a) Bi$_2$Se$_3$/Cu(111) and (b) Bi$_2$Se$_3$/Co(0001) bilayers. Panels (c) and (d) show charge rearrangement around the interface of bilayers in panels (a) and (b), respectively.} 
	\label{fig:fig5}
\end{figure*}

Finally, we propose a purely charge transport measurement that could detect which among the spin-textures shown in Figs.~\ref{fig:fig3}(b)--(d) resides at the Fermi level of TI/FM interface. Our scheme  requires to fabricate vertical heterostructure in Fig.~\ref{fig:fig1}(b) and measure its tunneling anisotropic magnetoresistance (TAMR). The TAMR is a phenomenon observed in magnetic tunnel junctions with a {\em single} FM layer~\cite{Mahfouzi2014a,Park2008,Chantis2007,Matos-Abiague2009}, where SOC makes the band structure anisotropic so that the  resistance of such junctions changes as the  magnetization $\mathbf{m}$ is rotated by angle $\theta$ or $\phi$ in Fig.~\ref{fig:fig1}(b). The resistance change is quantified by the TAMR ratio defined as~\cite{Chantis2007,Matos-Abiague2009}
\begin{equation}\label{eq:tamr}
\mathrm{TAMR}_{\mathrm{out}(\mathrm{in})}(\alpha) = \frac{R(\alpha) - R(0)}{R(0)}.
\end{equation}
Here $\alpha \equiv \theta$ for $\mathrm{TAMR}_\mathrm{out}$ where magnetization in Fig.~\ref{fig:fig1}(b) rotates in the plane perpendicular to the TI/FM interface, and $\alpha \equiv \phi$ for $\mathrm{TAMR}_\mathrm{in}$ where magnetization in Fig.~\ref{fig:fig1}(b) rotates within the plane of the TI/FM interface. In the case of $\mathrm{TAMR}_\mathrm{out}(\theta)$, $R(0)$ is the resistance when $\mathbf{m} \parallel \hat{z}$ in Fig.~\ref{fig:fig1}; and in the case of $\mathrm{TAMR}_\mathrm{in}(\phi)$, $R(0)$ is the resistance when $\mathbf{m} \parallel \hat{x}$ in Fig.~\ref{fig:fig1}. Thus, $\mathrm{TAMR}_\mathrm{out}(\theta)$ changes due to the different orientations of the magnetization with respect to the direction of the current flow, while the situation
becomes more subtle for  $\mathrm{TAMR}_\mathrm{in}(\phi)$ where the magnetization remains always perpendicular to the current flow. Figure~\ref{fig:fig4}(b) demonstrates that the largest $\mathrm{TAMR}_\mathrm{out}(\theta=\pm 90^\circ)$ is obtained by tuning the Fermi level to  \mbox{$E_F-E_F^0=-0.35$} eV so that nearly helical spin texture in Fig.~\ref{fig:fig3}(d) resides at the Fermi level. Another signature of its presence is rapid increase of $\mathrm{TAMR}_\mathrm{out}(\theta)$ when tilting $\mathbf{m}$ by small angles $\theta$ away from the current direction. The in-plane $\mathrm{TAMR}_\mathrm{in}(\phi)$ shown in the inset of Fig.~\ref{fig:fig4}(b) is much smaller (and difficult to converge in the number of transverse $k$-points) quantity which does not differentiate between spin textures shown in Figs.~\ref{fig:fig3}(b)--(d).

\section{Methods}

We employed the interface builder in the {\tt VNL}~\cite{vnl} and {\tt CellMatch}~\cite{Lazic2015} packages to construct a common unit cells for:  (a) Bi$_2$Se$_3$/Cu(111) bilayer, where the common unit cell is 5 $\times$ 5 in size compared to the smallest possible Cu(111) slab cell and copper is under compressive strain of 0.9 \% while Bi$_2$Se$_3$ lattice constant is unchanged; (b) Bi$_2$Se$_3$/Co(0001) bilayer where Co(0001) has the same lattice constant as Bi$_2$Se$_3$, so the same unit cell as for Cu(111) is used without any strain on Co(0001). These two unit cells are illustrated in Figs.~\ref{fig:fig5}(a) and ~\ref{fig:fig5}(b), respectively. In order to determine the best stacking of atomic layers and the distance of Bi$_2$Se$_3$ atoms with respect to surfaces of Cu(111) and Co(0001), we use DFT calculations  as implemented in the  {\tt VASP} package~\cite{Kresse1993}. The electron core interactions are described by the projector augmented wave (PAW) method~\cite{Blochl1994}, and vdW-DF~\cite{Dion2004} with optB88 is used as density functional~\cite{Mittendorfer2011} in order to describe van der Waals (vdW) forces between QLs of  Bi$_2$Se$_3$ or between Bi$_2$Se$_3$ and metallic layers. The cutoff energy for the plane wave basis set is 520 eV for all calculations, while $k$-points were sampled at 3$\times$3 surface mesh. We use Cu and Co layers consisting of 5 MLs, where 3 bottom MLs are fixed at bulk positions while the top two metallic MLs closest to Bi$_2$Se$_3$ are allowed to fully relax until forces on atoms drop below \mbox{$1$  meV/\AA}. In order to avoid interaction with periodic images of the bilayer, \mbox{$18$ \AA} of vacuum was added in the $z$--direction. 

For the case of Bi$_2$Se$_3$ on Co(0001), the most favorable position yields a binding energy of 460 meV per Co atom. Both ML of Co and QL of Bi$_2$Se$_3$ in direct contact gain some corrugation, roughly around \mbox{$\simeq 0.1$ \AA}, while the average $z$--distance between them is \mbox{2.15 \AA}. The average distance between the ML of Cu and QL of Bi$_2$Se$_3$ in direct contact is around \mbox{2.26 \AA} with smaller corrugation than in the case of Co(0001), while the binding energy is 294 meV per Cu atom. For other relative positions of Bi$_2$Se$_3$ layer with respect to Cu(111) and Co(0001) layers the difference in binding energy is very small. Binding energies in both cases are rather small, thereby signaling the dominant vdW forces. Nevertheless, some charge rearrangement does occur at the interface due to push back/pillow effect~\cite{Vazquez2007}, as shown in Figs.~\ref{fig:fig5}(c) and ~\ref{fig:fig5}(d) where charge rearrangement is more pronounced for the case of Bi$_2$Se$_3$/Cu(111) interface.

The calculation of the retarded GF in Eqs.~\eqref{eq:oneleadgf} and ~\eqref{eq:twoleadgf} requires $\mathbf{H}^\mathrm{DFT}_{\mathbf{k}_\parallel}$ represented in the linear
combination of atomic orbitals (LCAO) basis set which makes it possible to spatially separate system into the active region attached to one or two semi-infinite leads, as illustrated in Figs.~\ref{fig:fig1}(a) and ~\ref{fig:fig1}(b), respectively. We employ {\tt ATK} package~\cite{atk} for pseudopotential-based LCAO noncollinear DFT calculations yielding  $\mathbf{H}^\mathrm{DFT}_{\mathbf{k}_\parallel}$, from which we obtain retarded GFs and the corresponding spectral functions, as well as the resistance in Eq.~\eqref{eq:resistance}. In {\tt ATK}  calculations, we use Perdew-Burke-Ernzerhof (PBE) parametrization of generalized gradient approximation for the exchange-correlation functional; norm-conserving pseudopotentials for describing  electron-core interactions; and LCAO basis set generated by the {\tt OpenMX} package~\cite{Ozaki2003,openmx} which consists of {\tt s2p2d1}  orbitals  on  Co, Cu and Se atoms, and {\tt s2p2d2} on Bi atoms. These pseudoatomic orbitals were generated by a confinement scheme~\cite{Ozaki2003} with the cutoff radius 7.0 and 8.0 a.u. for  Se  and  Bi  atoms,  respectively, and 6.0 a.u. for Co and Cu atoms. The energy mesh cutoff for the real-space grid is chosen as 75.0 Hartree. 

\section{Acknowledgments}
We thank Takafumi Sato and Jia Zhang for insightful discussions. J. M. M.-T., K. D. and B. K. N. were supported by NSF Grant No. ECCS 1509094. J. M. M.-T. also acknowledges support from Colciencias (Departamento Administrativo de Ciencia, Tecnologia e Innovacion) of Colombia. P. L. was supported by the Unity Through Knowledge Fund, Contract No. 22/15 and H2020 CSA Twinning Project No. 692194, RBI-T-WINNING. S. Smidstrup, D. Stradi and K. Stokbro acknowledge support from the European Commission Seventh Framework Programme Grant Agreement III–V-MOS, Project No. 61932; and Horizon 2020 research and innovation programme under grant agreement SPICE, project No 713481.  The supercomputing time was provided by XSEDE, which is supported by NSF Grant No. ACI-1053575. 






\end{document}